\documentclass{elsarticle}
\usepackage{hyperref}
\journal{Carbon}

\makeatletter
\def\ps@pprintTitle{%
  \let\@oddhead\@empty
  \let\@evenhead\@empty
  \let\@oddfoot\@empty
  \let\@evenfoot\@oddfoot
}
\makeatother

\usepackage{xcolor, soulutf8}
\sethlcolor{yellow}

\usepackage{float}
\restylefloat{figure}
\usepackage{graphicx}
\usepackage{dcolumn}
\usepackage{bm}
\usepackage{inputenc}
\usepackage{multirow}
\usepackage{upgreek} 
\usepackage{amsmath}
\usepackage{makecell}
\usepackage[normalem]{ulem}

\hypersetup{
colorlinks=true,
linkcolor=blue,
citecolor=blue,
urlcolor=blue
}
\graphicspath{{./immagini/}}
\begin{document}
\begin{frontmatter}
\title{Adhesion, Friction and Tribochemical Reactions at the Diamond-Silica Interface}
\author{Michele Cutini}
\author{Gaia Forghieri}
\author{Mauro Ferrario}
\author{Maria Clelia Righi\corref{mycorrespondingauthor}}
\address{Department of Physics and Astronomy, University of Bologna, 40127 Bologna, Italy}
\ead{clelia.righi@unibo.it}
\cortext[mycorrespondingauthor]{Corresponding author}

\begin{abstract}
Diamond-based coatings are employed in several technological applications, for their outstanding mechanical properties, biocompatibility, and chemical stability. Of significant relevance is the interface with silicon oxide, where phenomena of adhesion, friction, and wear can affect drastically the performance of the coating. Here we monitor such phenomena in real-time by performing massive \emph{ab initio} molecular dynamics simulations in tribological conditions. We take into account many relevant factors that can play a role, i.e. the diamond surface orientation and reconstruction, silanol density, as well as, the type and concentration of passivating species. The large systems size and the long simulations time, put our work at the frontier of what can be currently done with fully \emph{ab initio} molecular dynamics. The results of our work point to full hydrogenation as an effective way to reduce both friction and wear for all diamond surfaces, while graphitization is competitive only on the (111) surface. Overall we expect that our observations will be useful to improve technological applications where the silica-diamond interface plays a key role. Moreover, we demonstrate that realistic and accurate \emph{in silico} experiments are feasible nowadays exploiting HPC resources and HPC optimized software, paving the way to a more general understanding of the relationship between surface chemistry and nanoscale-tribology.
\end{abstract}

\end{frontmatter}

\section{Introduction}
Diamond is a material with widespread potential applications, ranging from electronics, optics and biocompatible devices~\cite{WORT200822,Lagomarsino2015}
Diamond and diamond-like carbon (DLC) 
are also excellent low-friction coatings~\cite{rev-dlc,Erdemir2006} employed in many industrial 
applications to reduce friction and protect the surfaces they cover from wear. However, the performances of diamond-based coatings depend heavily on 
environmental conditions, which refrains the extensive adoption of diamond materials in tribology. 
For instance, diamond lubrication is poor in vacuum or dry atmospheres compared to hydrogen 
and water-rich ones~\cite{Kumar2011,Konicek2012,DeBarrosBouchet2012,Konicek2008}. 
The presence of sacrificial molecules with affinity to the carbon surface is crucial to obtain 
surface passivation and thus low friction in carbon films~\cite{Konicek2008, PhysRevLett.111.146101}. 
Indeed, experiments and simulations agree to indicate carbon dangling bonds as sources of friction 
in diamond-based materials~\cite{Qi2006,Dag2004,Zilibotti2009}. 
Wear rate is related to counter-surface chemical nature~\cite{Aboua2020,Aboua2018} and diamond surface
chemistry. Surfaces with high content of $sp^2$ graphite-like bonding network are believed to have
wear-protection effect~\cite{Liu1996}.  In this complex scenario a clear understanding of basic
chemical events responsible for diamond-based materials' frictional and wear performances seems
indispensable to improve the performances of this class of materials to wide operational conditions.
\newline
Diamond is also an excellent material for cutting and polishing applications due to its extreme 
hardness and chemical inertness. Counter-intuitively some softer materials can polish a diamond 
surface~\cite{Thomas2014,Thomas2014a}, and they are used to produce defect- and damage-free 
diamond films. This polishing step is crucial for the functioning of diamond-based devices~\cite{Chen2017}. 
An emerging technique to polish diamond is the chemical-mechanical polishing (CMP)~\cite{Xiao-rev2022} 
with silica particles. This procedure has the advantages of operating at low temperatures and employing 
materials already in integrated circuit industry supplies. Efforts were devoted to understanding the 
chemical processes underlying silica-drive CMP. One mechanism proposed~\cite{diamond-wear1,Luo2021} 
for the C(110) surface sees silica pilot atoms pulling and breaking C--C bonds. 
Semi-empirical Molecular Dynamics (MD) simulations on minimal diamond-silica models indicate that both Si--O--C and Si--C bonds 
can promote C--C bond-breaking events, but only when adatoms are adsorbed vicinal to the reaction
center~\cite{Peguiron2016}. It is still unclear how surface termination and orientation can affect the tribological features of this interface. Recently, classical MD simulations provided a qualitative description  of diamond CMP with silica particles 
in the presence of oxidizing species~\cite{Kawaguchi2021,Guo2019}. Regardless of the efforts, 
the comprehension of silica-driven diamond wear remains at a preliminary stage.
\newline
In this work, we have studied the diamond-silica interface in tribological conditions using 
large-scale \emph{ab-initio} Molecular Dynamics (AIMD) simulations. 
The diamond-silica interface is not only relevant for technological applications, but it is 
also a very interesting test case for the understanding of friction and wear mechanisms 
at the nano-scale~\cite{Milne2019}. Our major aim is to identify the critical chemical processes 
that are triggered when the systems evolves under tribological conditions. To draw a picture for 
a realistic case, we have systematically analyzed the effect of surface facets and chemical groups 
commonly produced at the interface between diamond and silica. 
Micro- and nano-crystalline diamond coatings typically expose low-index facets~\cite{SCHADE20061682}, 
which have different chemical reactivity and $sp^2$  reconstruction. So we have considered the 
most common, different, diamond facets, i.e. reconstructed C(111), C(001), C(110), together with reactive 
under-coordinated surfaces to better assess the role of dangling bonds. We have also investigated 
the effects of full and partial passivation involving the molecular species typically present in 
the atmosphere, such as hydrogen, water, and oxygen, elucidating the kinetics of passivation in 
ideal and tribological conditions. For silica we have analyzed different silanol densities, 
which vary drastically with sample preparation. In all cases, we have monitored the resistive force 
arising in response to sliding during the simulations under tribological conditions, to search for 
direct correlations between chemical events and friction properties. Our results permit to 
identify the most efficient features of the surfaces which can enhance lubrication and wear protection
at this fundamental and technologically relevant materials interface. 

\section{Computational Methods}
We performed static Density Functional Theory (DFT)~\cite{DFT} calculations by means of the PW-scf 
package \textsc{Quantum~ESPRESSO} suite~\cite{Giannozzi-2009,Giannozzi-2017,Giannozzi-2020}. 
For the AIMD simulations of sliding surfaces we used a modified version by our group.
Spin polarized calculations are performed to take into account the possible magnetization arising 
from the presence of surface dangling bonds and bond-forming breaking events. The electronic wave 
function is expanded on a plane-wave basis set, using ultrasoft pseudopotentials to describe core 
electrons~\cite{Vanderbilt-1990}.  We employed the generalized gradient approximation (GGA) within 
the Perdew-Burke-Ernzerhof (PBE) parameterizations as exchange-correlation functional~\cite{PBE-1996}, 
including London dispersion forces by adopting the D2 scheme\cite{Grimme-D2-2006}. This computational approach proved to give accurate estimations of organic \cite{Cutini2019,Cutini2019b,Cutini2021}, inorganic \cite{Cutini2019a}, and composites\cite{Cutini2017a} materials properties. The kinetic energy 
cutoff to expand the wave function (ecutwfc) and the charge densities (ecutrho) are 30~Ry and 240~Ry, 
respectively. We ensured full convergence of the Brillouin zone sampling by using $2\times 2\times 1$ 
and $2\times 1\times 1$~grids for cells with (a,b) lattice parameters equal to (10.11~\AA,~8.76 \AA) 
and (10.11~\AA,~17.89 \AA), respectively.
Molecular dynamics simulations are performed on the electronic ground state with an \emph{ad hoc} 
modifications of the code to impose tribological conditions. AIMD simulations are run considering 
only the $\Gamma$ point for reducing their high computational burden  caused by the large models sizes 
and performing the integration 
in time by means of Verlet algorithm with a timestep $\delta t = 30\,\text{a.u.} \simeq 1.45\,
\text{fs}$ for a typical production time of $15\,\text{ps}$. 
The sliding under pressure of silica on a diamond substrate was driven by constraining the velocity 
in the $x$--direction of the outmost Si atoms of silica to a constant value of $200\, m/s$, 
while applying onto them external forces in the $z$--direction equivalent to a normal load of $1\,
\text{GPa}$ and freezing the dynamics of the basal C atoms of the diamond substrate. 
Temperature was controlled by setting a reference value of $300\,\text{K}$, initially by 
sampling the Maxwellian distribution and along the dynamical trajectory by applying separate 
thermostats to silica and diamond, while custom calculating the kinetic energies to be compared 
to the requested target temperature and adjusting only the thermal component of the particle’s 
velocities. 
The trajectories generated by the AIMD simulations were further analyzed by visual inspection and 
by monitoring the interfacial distance $\Delta z$ between the silica and diamond surfaces, obtained  
subtracting the average $z$ coordinates of the interfacial Si and C atoms and the components of 
what we hereby call the resistive force, $\mathbf{F}$, i.e. the force opposed to the motion 
of silica, which is a direct indicator of the friction generated at the silica-diamond interface. 
The value of $\mathbf{F}$ is evaluated at every dynamical steps as:

\begin{equation}
    \mathbf{F} = \sum_{Si}\mathbf{f}^{(Si)}\, ,
    \label{eq:mu}
\end{equation}
where $Si$ are the silicon atoms onto which the external load forces and velocity constraints 
were applied. Reported errors on $\mathbf{F}$ and $\Delta z$ were calculated by carrying out 
the block average procedure described in details by Templeton et al~\cite{Templeton2021} and references therein.

To understand the effect of tribological conditions on the chemical processes involving the passivating molecules, 
we computed the chemisorption energy barriers of H$_{2}$ and O$_{2}$ and H$_{2}$O on the diamond
surfaces, using the the Nudged Elastic Band (NEB) method \cite{NEB:jonsson1998, NEB:henkelman2000-1}. 
For NEB calculations, we set the elastic constants to $1.0\,a.u.$, using 10 images to link initial 
and final configurations and an energy threshold of $0.05\, a.u.$.
\newline
Amorphous silica surface was built starting from a crystalline supercell of alpha quartz, thermally 
annealed using a classical force field~\cite{Vashishta-1990}, and then cut to create a thin film. 
We removed all reactive dangling bonds on the surfaces by adding H atoms to Si--O$\cdot$ and 
HO$\cdot$ groups to Si$\cdot$. We employed supercells with ad-hoc lateral dimension chosen 
to match the amorphous silica with the diamond substrate, creating large 2D-periodic models. 
Models built with the (111) diamond facet have between 150 and 200 atoms, while those built 
with (110) and (001) facets have between 320 and 380 atoms. To avoid spurious interactions, 
at least 15~\AA~of vacuum was added in the vertical direction in the PW-scf calculations 
for all surfaces and interfaces models. The large number of electrons and plane waves implied 
by the use of these models makes the calculation of AIMD trajectories computationally demanding. 
Indeed, 63000 GPU hours were employed to obtain the results here presented.

\section{Surface Models}
\subsection{Modeling of the Amorphous Silica Surface}

\begin{figure*}
    \centering
    \includegraphics[width=0.98\textwidth]{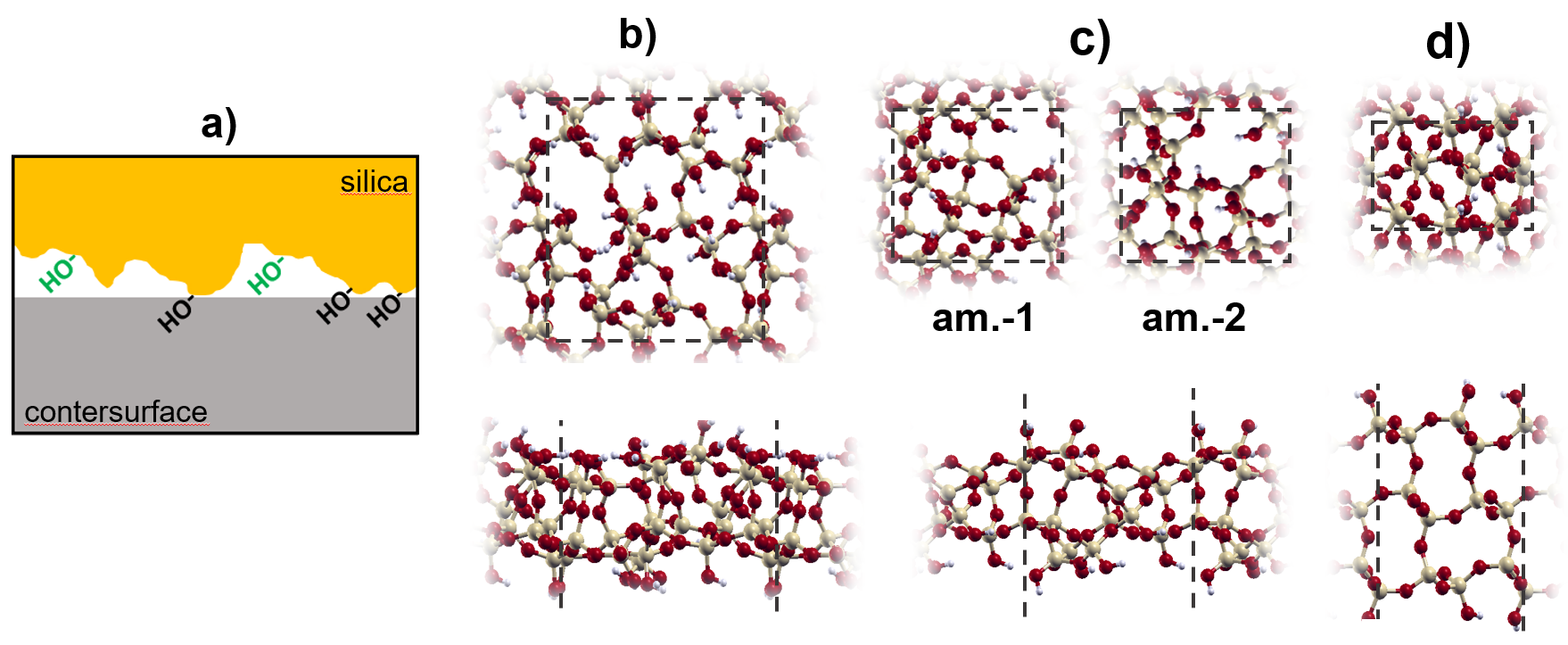}
    \caption{ (a) Schematic representation of the interface between a silica substrate and a countersurface. 
    Inner (exposed) silanol groups are reported in green (black). Top (first row) and side (second row) views of the silica surface models: 
    (b) from Ugliengo et al.~\cite{Ugliengo2008}, (c) from this work, amorphous 1 (am.1) and amorphous 2 (am.2), which differ from silanols density, (d) for the 
    cristobalite (101) surface.}
    \label{fig:silica-models}
\end{figure*}

In many technological applications of silicon is in contact with air and present a surface layer of silicon oxide, i.e., silica, which has an amorphous structure. The amorphous surface 
exposes siloxane bridges (Si--O--Si) and silanol groups (Si--OH). Different types of silanol groups
can be distinguished depending on their chemical environment and reactivity, e.g. isolated, vicinals, 
interacting and geminal~\cite{Rimola2013}. The location of silanols on the rough amorphous surface plays 
an important role in tribology. Indeed, exposed silanols can interact with a countersurface, 
while the inner ones are protected from the tribo-reactions, as pictured in the (a) panel of 
Fig.~\ref{fig:silica-models}.

The lack of long-range periodicity and the presence of a large number of superficial features poses a major 
challenge in accurately modelling the amorphous silica surface. Several efforts were devoted in the past to 
develop an accurate model for \emph{ab initio} studies~\cite{Ugliengo2008, Ewing2014}. 
In Tab.~\ref{tab:silica-models} we have summarized the main structural features of the model prepared 
by Ugliengo et al.~\cite{Ugliengo2008,Gignone2015}. The model correctly reproduces the average density 
of silanols observed on amorphous silica samples prepared with different methods 
(4.8 OH/nm\textsuperscript{2})~\cite{CROISSANT2018181}, and exhibits a balanced number of exposed 
and inner silanols. The main drawback is represented by its wide surface area, as reported in the last column of 
Tab.~\ref{tab:silica-models}. The wider the surface, the higher the atoms number needed to build a 
silica--diamond interface, and so the computational cost. Among the crystalline SiO\textsubscript{2} 
polymorphs, cristobalite exhibits, on the (101) surface, a superficial silanol density in line with 
the average experimental value over a relatively small reference surface area~\cite{Musso2009}, 
see panel (d) of Fig. \ref{fig:silica-models}. Unfortunately, due to its crystalline nature, the 
Si--O--Si angle distribution is narrow and silanol groups are exposed, which over-enhances the surface 
tribological reactivity. In this scenario, we have decided to consider an amorphous silica model, 
to represent in a more realistic way the silicon oxide surface than by using cristobalite, 
but with a reduced later dimension with respect the amorphous model of Ref.~\ref{tab:silica-models} 
to make the AIMD simulations of the silica-diamond interface computationally feasible. 
With our modelling procedure based on the thermal annealing of the bulk structure, described in details 
on the previous section, we obtained two amorphous silica slab with the desirable surface characteristics.
The surfaces have silanol density slightly higher and slightly lower of the average experimental value, 
modelling silanol-rich and -poor surface regions. The number of exposed and inner silanols is balanced, 
the surface area is limited, and the wide distribution of the Si-O-Si angle indicates a disordered 
structure, in agreement with the amorphous nature of the surface.  Numerical values are reported in 
Tab~ \ref{tab:silica-models} and a view of the exposed surfaces is shown in the (c) panel 
of Fig.~\ref{fig:silica-models}. We would like to point out that our procedure allowed 
to produce silica models which are more realistic representations of silica surface compared 
to models present in literature based on un-reconstructed crystalline surface exhibiting 
reactive Si$\cdot$ and O$\cdot$ dangling bonds \cite{Peguiron2016}.
\begin{table*}
\centering
\caption{\small Structural parameters of the amorphous and crystalline silica surface models.
The average values and standard deviations of the Si--O--Si angle, the numbers of inner and exposed 
silanol (Si--OH) groups per cell, as well as the overall surface density of silanols are reported. 
As an indication of the model size, we have included the cell area $\mathbf{A}$.}
\begin{tabular}{l  c c c c c}
& \thead{$\mathbf{SiOSi}$ \\ (\textbf{°})} & \thead{$\mathbf{n.SiOH}$ \\ \textbf{inner}} & \thead{$\mathbf{n. SiOH}$ \\ \textbf{exposed}} &
\thead{$\mathbf{\rho - SiOH}$ \\ (\textbf{nm$^{-2}$})} &
\thead{$\mathbf{A}$ \\ (\textbf{nm$^2$})}  \\
\hline
Amorphous from Ref.\cite{Ugliengo2008} & 144$\pm$11 & 2 & 6 & 4.8 & 1.60 \\
Cristobalite (101) & 146$\pm$4  & 0 & 2 & 4.6 & 0.76 \\
Amorphous-1 (this work)  & 141$\pm$15 & 1 & 2 & 3.4 & 0.89 \\
Amorphous-2 (this work)  & 141$\pm$15 & 3 & 2 & 5.6 & 0.89 \\
\end{tabular}
\label{tab:silica-models}
\end{table*}
\begin{verbatim}
\end{verbatim}

\subsection{A Realistic Diamond Surface}

A realistic diamond surface displays different 
crystal faces, reconstructed regions, total-or-partial passivated areas, and defects, such as reactive 
carbon atoms with dangling bonds. To rationalize the role in tribology of each of these features we performed 
a systematic analysis on a set of ideal diamond surfaces representative of all the cases mentioned above. 
In detail, we have considered: i) three diamond surfaces with low Miller indices, which 
are expected to mainly characterize a realistic diamond surface, i.e., C(111), C(110), and C(001), where
for the (111) facet we considered both the  un-reconstructed C$\cdot$(111) and Pandey-reconstructed configurations; 
ii) the C(111) surface passivated by three common atmospheric molecules, e.g., water, oxygen, and hydrogen;  
iii) different levels of passivation, i.e. 0\%, 50\% and 100\% for C(111), 25\% and 50\% for C(110), 50\% for C(001), 
choosing hydrogen as the passivating specie for its high affinity to carbon.

\section{Results and Discussion}
\subsection{Tribological Simulations}

\begin{figure*}[!h]
    \label{fig:optgeom}
    \centering
    \includegraphics[width=0.98\textwidth]{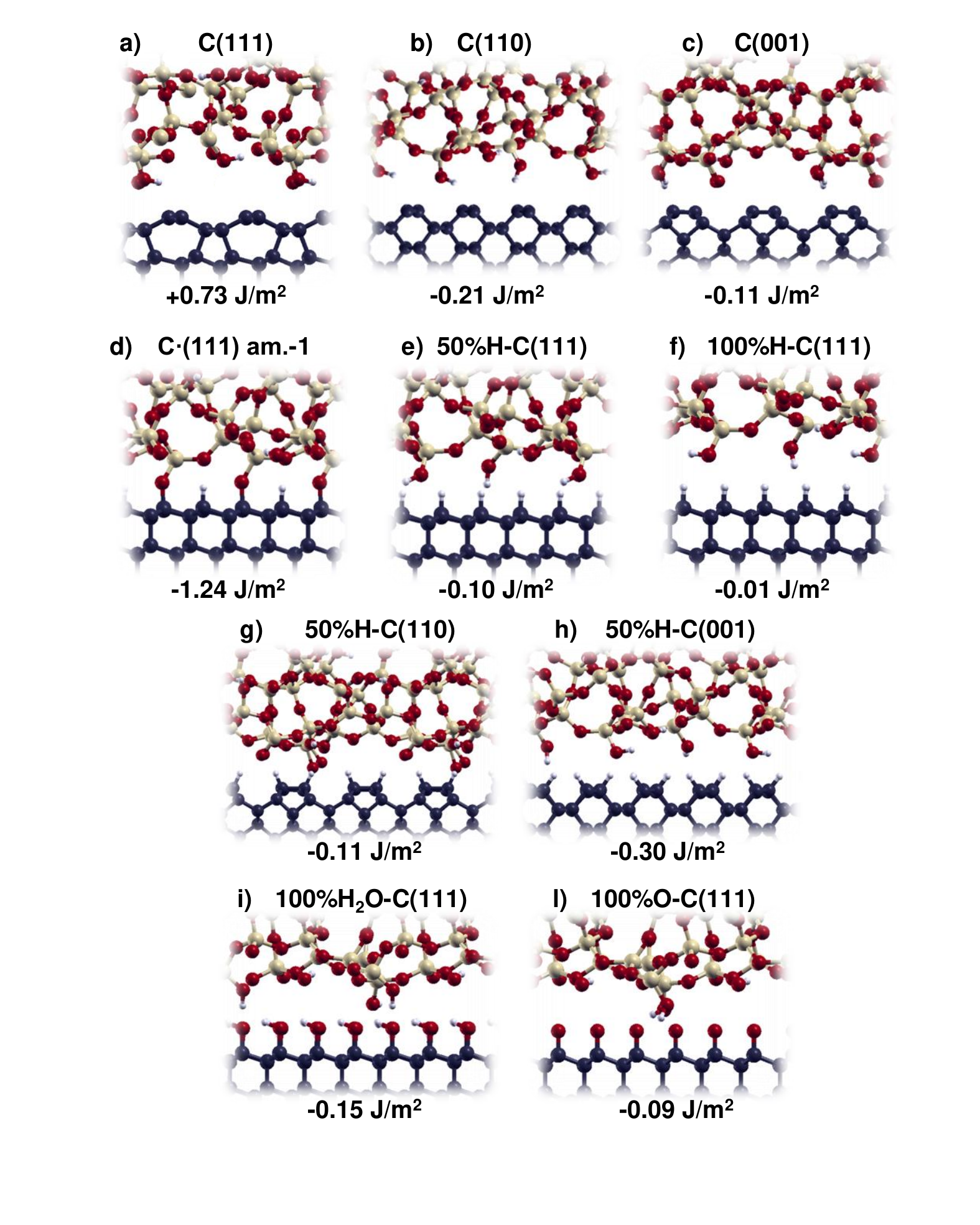}
    \caption{Silica-diamond interfaces considered in this work and computed $E_{ADH}$. Each structure is optimized under 1GPa of normal load.}
\end{figure*}

This section focuses on the tribological properties of the twelve interfaces obtained by assembling 
the diamond and silica models mentioned in the previous section. All diamond surfaces are mated with 
the silica amorphous-1 model which has a 3.4 OH/nm\textsuperscript{2} density. The C$\cdot$(111) surface 
is matched also with the silica amorphous-2 model which has a 5.6 OH/nm\textsuperscript{2} density 
for assessing the effect of silanol density. After relaxing the structures under 1 GPa of load no 
bond-breaking-forming events is observed, but in one case, i.e. for the C$\cdot$(111) surface. 
This diamond surface exposes reactive carbon dangling bonds (C$\cdot$), which cause the breaking of 
a silica silanol group (Si-O-H) to form a C-H bond on the diamond surface and a Si-O-C bridge across 
the interface. Chemical bonds connecting the substrate and the counter-surface are highly relevant 
for tribological applications being sources of adhesive friction. 
Atomistic views of the optimized interfaces are reported in Fig.~\ref{fig:optgeom} 
along with the computed adhesion energy ($E_{ADH}$) which is defined as the  difference between the energy of the optimized interface under 1 GPa of load and the energy of the two isolated relaxed surfaces. 
In most cases, the $E_{ADH}$ span from -0.01 J/m\textsuperscript{2} to -0.30 J/m\textsuperscript{2} indicating 
that the surfaces interact through weak non-covalent interaction. 
The high $E_{ADH}$ value for the C$\cdot$(111) am.-1 case is a clear evidence of chemical bond formation across the interface. 
Instead, the peculiar positive value of $E_{ADH}$ for the C(111) shows that normal load can perturb 
the graphite-like surface bonds on the (111) diamond facet destabilizing the interface.

\begin{figure*}
    \centering
    \includegraphics[width=0.98\textwidth]{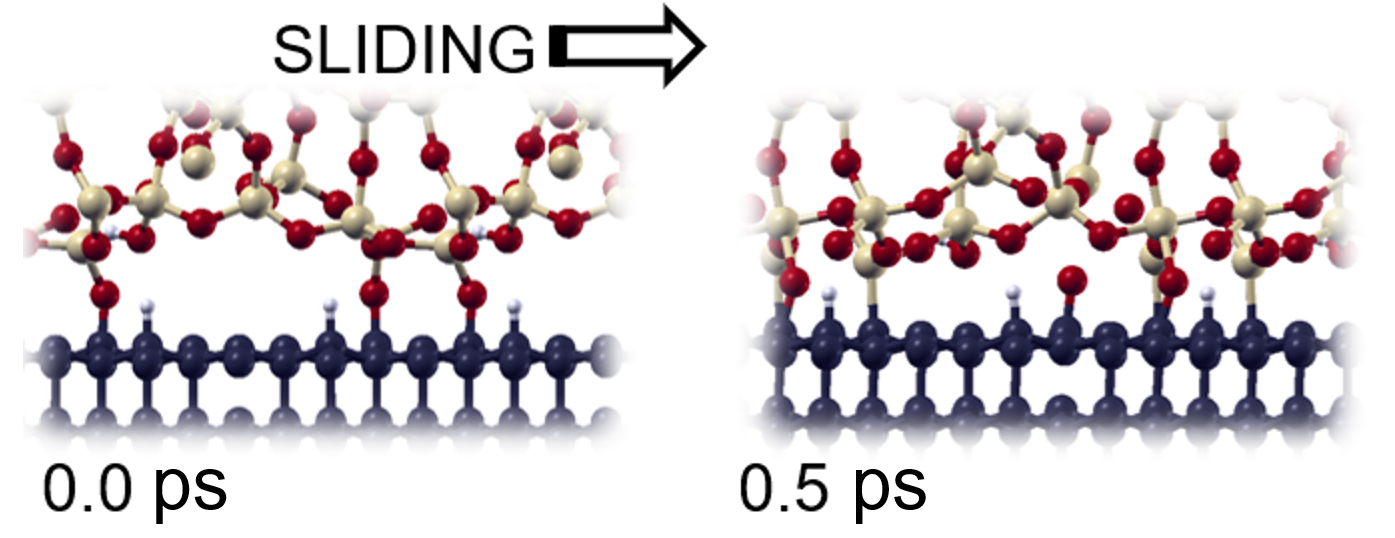}
    \caption{Selection of chemical events occurring during the tribological simulations 
    for the interface between silica and C$\cdot$(111) diamond surface.}
    \label{fig:movies}
\end{figure*}

We have then simulated  using AIMD the sliding of these interfaces under load. The chemical events observed 
during the dynamics depend significantly on the features of the diamond surface. Dangling bonds work as 
reactive centers driving the breaking of silica silanols with the consequent formation of Si--O--C bridges 
as previously described. The silica sliding leads these interfacial bonds to stretch and eventually to break 
with the formation, in the case of the C(111) surface, of C--O$\cdot$ groups on the diamond surface and 
Si--C bridges across the interface, see Fig. \ref{fig:movies} and supplementary material.
The presence of partial hydrogen passivation reduces diamond's reactivity, limiting the formation of 
Si--O--C and Si--C interfacial bonds. Bond-breaking and bond-forming events are entirely inhibited 
by full passivation, independently from the passivating species.
In line with our findings, semiempirical simulations on DLC coatings indicate that by increasing the hydrogen content in the carbon film the running-in period is reduced and the cold sealing between the substrate and the countersurface is prevented \cite{Pastewka2010}.
Interestingly, the presence of C=C bonds on the diamond surface has a different effect depending on the exposed 
crystal face. Remarkably, the Pandey reconstruction of the C(111) surface remains inert for the whole 
simulation time. Instead, the C(110) and C(001) weaker superficial C=C bonds break when they interact with 
silanols, forming Si--O--C and Si--C interfacial bridges. Furthermore, we observe wear in one case, 
for C(110) diamond surface, in agreement with Peguiron et al.~\cite{Peguiron2016}. 

During the first picoseconds of dynamics a series of chemical events driven by silica lead a diamond C–C bond to break permanently, see supplementary material. In Fig.~\ref{fig:legami} a detailed analysis of the chemical bond lengths involved in the wear of the diamond is reported along with the oscillation of the interface potential energy. We observe in order: i) the chemical attack by a Si-O$\cdot$ group on the diamond surface, Fig.~\ref{fig:legami} panel a)-b), which leads to the formation of a Si-O-C\textsubscript{1} bond (see green line in Fig. \ref{fig:legami}). This bond formation stabilizes the interface so reducing the $\Delta$E value (blue line). ii) The silica motion stretches the Si-O-C\textsubscript{1} bond, shortening the Si-C\textsubscript{3} distance, eventually leading to the formation of the Si-C\textsubscript{3} bond (red line). At this point of the reaction, both ad-atoms (O and Si) reside on zig-C atoms (C\textsubscript{1} and C\textsubscript{3}). This adsorption geometry is known to destabilize the C-C bonds between the superficial zig-zag carbon chains and the underlying diamond bulk atoms \cite{Peguiron2016},  i.e., C\textsubscript{1}-C\textsubscript{B} and C\textsubscript{3}-C\textsubscript{B}. Indeed, as the Si-C\textsubscript{3} bond is fully established, the C\textsubscript{1}-C\textsubscript{B} bond starts breaking (black line) with the simultaneous shortening of the C\textsubscript{1}-O$\cdot$ bond indicating the formation of a C=O double bond (panel c), which remains stable
for the whole simulation time. iii) The countersurface sliding leads to the C\textsubscript{3}-C\textsubscript{B} bond breaking at 4.3 ps, see panel d and yellow line. At this point of the simulation, the chemical bonds are highly stretched as indicated by the peaking of the potential energy. iv) Higher stability is obtained when the Si-C\textsubscript{3} bond breaks and the underlying C\textsubscript{3}-C\textsubscript{B} bond is reformed (panel e).

We can speculate that the formation of a C=O double bond is a precursor of $CO_2$ as a product
of diamond degradation. Indeed, $CO_2$ is a common wear product of diamond CMP. During this reaction chains 
also silica wears, but differently from diamond, it can quickly regenerate by hydration. These events may 
explain the unusual phenomenon of diamond wearing obtained with softer oxides. Interestingly, hydrogen 
passivation reduces wear. At 25\% and 50\% hydrogen passivation, diamond C-C bond breaking events are less common and reversible. 

The H-passivation reduces the number of superficial C atoms available for bonding to Si$\cdot$ radicals promoting the attack of Si$\cdot$ to C=O. The resulting Si-O-C bond leads to the re-establishment of the C-C bond with the bulk. Furthermore, when the two C atoms vicinal to a C=O group become tetrahedrally coordinated, they force the C=O group to lay parallel to the surface causing steric hindrance with the neighboring zig-zag chain of C. Consequently, it is more energetically favored to have the C=O double bond broken and the C-C bond with the bulk reformed. The partial H-passivation increases the number of superficial C with tetrahedral coordination thus making this reverse mechanism more likely to occur.

We expect similar results by replacing silica with other components of CMP slurries, such as Al$_{2}$O$_{3}$ and CeO$_{2}$, for their chemical similarity with SiO$_{2}$.\cite{slurries}

\begin{figure*}
    \centering
    \includegraphics[width=0.98\textwidth]{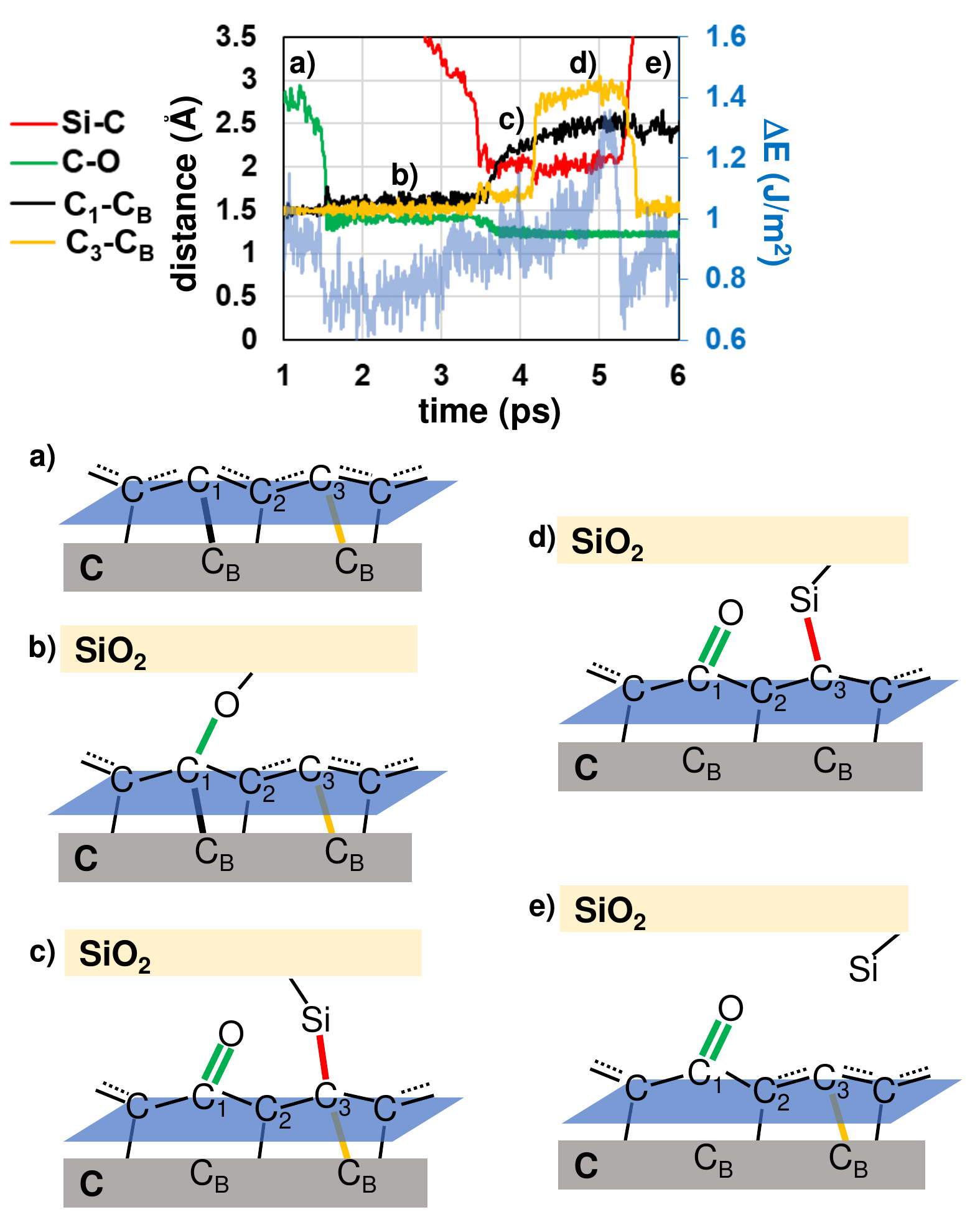}
    \caption{Time evolution of bond distance (see legend) and potential energy, $\Delta$E (cyan line), of the silica-driven wear of C(110) surface. $\Delta$E is defined as the difference between the energy of the i\textsuperscript{th} time step and the energy of the static relaxed interface. Distance$\slash\Delta$E refers to the left right vertical axes. The chemical events are graphically schematized in panels a)-e) for an easier understanding of the figure.}
    \label{fig:legami}
\end{figure*}

\begin{figure*}
    \centering
    \includegraphics[width=0.98\textwidth]{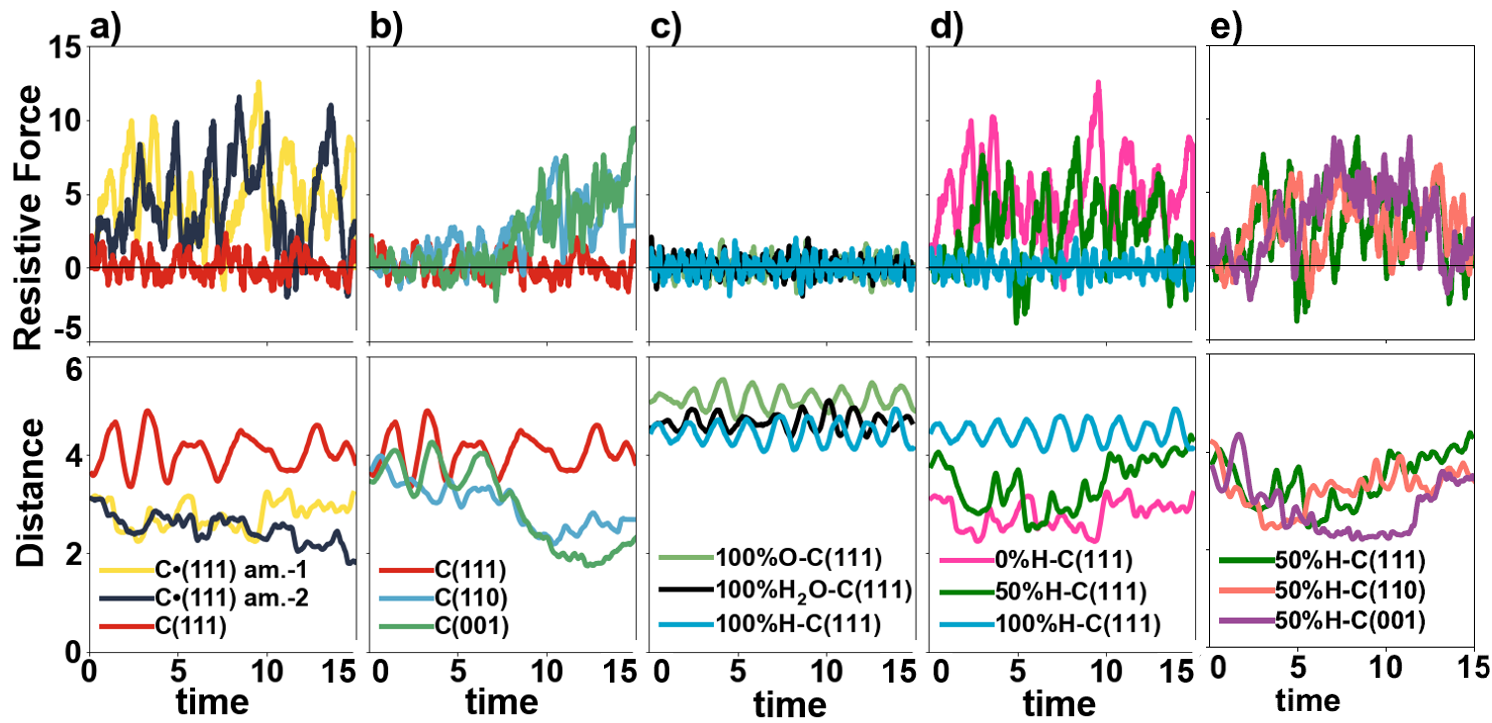}
    \caption{Time evolution (in ps) of resistive forces per area (in GPa) and interfacial distances (in  \AA) extracted 
    from the tribological simulations.}
    \label{fig:dyn}
\end{figure*}

The qualitative analysis of the AIMD trajectories strictly relates to the tribological figures of merit 
extracted from the dynamics. In Fig.~\ref{fig:dyn} we show the time evolution of the resistive force 
along the sliding direction and the interfacial distance during the tribological simulations. 
To filter out fast oscillations and improve the signal to noise ratio, moving averages over 200 timesteps (~0.3 ps) 
are plotted, rather than instantaneous values. 
As can be seen in panel a), where the tribological behavior of the interfaces of Figs. 2 a)-d) 
is compared with that of the interface between C$\cdot$(111) and silica amorphous-2, 
the resistive force is higher in the presence of many interfacial 
Si--O--C and Si--C bonds. These are easily formed on the C$\cdot$(111) surface regardless of silica silanol 
density (3.4 or 5.6 OH/nm$^2$). Removing defects on the C(111) surface by reconstruction or inhibits the 
reactivity, preventing bond formation and lowering the resistive force opposed to sliding. 
The interfacial distance reflects the nature of the interaction between substrate and countersurface. 
Short distances suggest chemical bond formation, large distances indicate weak non-covalent interactions. 

We observe similar tribological reactivity for both isolated and vicinal silanols. 
Geminal silanols, in which two hydroxyls group bond to a single Si atom, may behave differently. 
Nevertheless, they are by far the least common kind of silanols on amorphous silica surfaces~\cite{silica-new}. 
Therefore, we expect them to play a marginal role in the silica tribological features.
The more stable the diamond surface reconstruction, the longer it resists silica's chemical attack. 
In panel (b) of Fig.~\ref{fig:dyn} the effects of C=C double bonds on the surface is analayzed by comparing 
the tribological behaviour of the interfaces represented in Figs. 2 a)-c). 
One can see that higher peaks develop in the resistive force together with 
correlated drops in the interfacial distance signalling that while it takes about 5 ps for the C(110) 
surface to fail and longer, ~10 ps, for the C(001) one, C(111) stays inert, as already pointed out, 
for the whole simulation time. 
The rate of interfacial bond formation agrees with the known order 
C(110) $>$ C(001) $>$ C(111) for the diamond surface reactivity~\cite{Stekolnikov2002, DeLaPierre2014}. 
Even if we expect that, in the long run, some interfacial bonds will form eventually, our results 
indicate that the C(111) Pandey reconstruction is among the $sp^2$ type diamond termination the best 
suited for lubrication and wear prevention. 

In panel c) the effects of different passivating species is considered by comparing the interfaces of Fig. 2 f), i) and l).
Full chemical passivation inhibits diamond reactivity, thus the resistant force is lower and the interface 
is more distanced. Partial passivation has intermediate effects, reducing the number of reactive defect sites, 
but still allowing for the formation of a relevant number of interfacial bonds associated with relatively high values 
of the resistive force and low values of the interfacial distance, see panels (D) and (E) of Fig.~\ref{fig:dyn}.

\begin{figure*}
    \centering
    \includegraphics[width=0.98\textwidth]{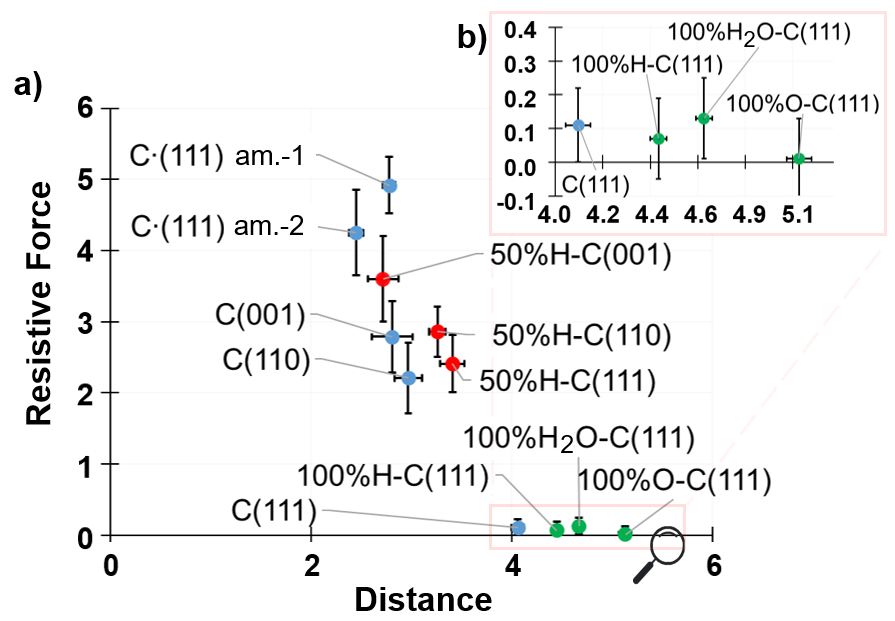}
    \caption{a) Summary of the mean resistant force (in GPa) and interfacial distance (in  \AA) values, b) zoom in of a figure portion. }
    \label{fig:summary}
\end{figure*}

A graphical summary is presented in Fig.~\ref{fig:summary} panel a) of the averages over the $15\,\text{ps}$ 
duration of the AIMD simulations for the component along the sliding direction of the resistive force 
and the interfacial vertical distance. The most effective lubrication of the C(111)diamond-silica interface 
occurs by oxygen chemical passivation, see Fig.~\ref{fig:summary} panel b). 
The O$_{2}$ passivated C(111) surface exposes C--O$\cdot$ groups 
that establish only weak hydrogen-bonds with silica silanols. Moreover, the oxygen coverage leads to an 
accumulation of negative charges on the diamond surface, which repulses oxygen-rich counter-surfaces 
to non-bonding distances. The effects of oxydation on the C(001) and C(110), 
which is explored in a forthcoming dedicated article, is different. 
For the unreconstructed C(111) surface, hydrogen passivation is as efficient as oxygen passivation. In this case, 
mainly weak dispersive forces drive interfacial interactions. The interfacial distance is shorter than 
for the oxygen passivation case due to the lower steric hindrance of hydrogen. Water passivation, instead, is slightly less efficient as the hydroxylated diamond surface establishes hydrogen bonds with silanols, with a slight increase of friction.
The results of a previous study \cite{ma15093247} on the tribological properties of amorphous carbons passivated with different species agree with our findings: -OH terminations generate higher friction than -H terminations.

The C=C termination is preserved only at the C(111) surface making it inert and preventing the formation of 
chemical bonds across the interface. For this case, in fact, the calculated force of resistance is of the 
same order of the one measured for the H$_{2}$ and O$_{2}$ passivated diamond surfaces, with a further 
reduced interfacial distance due to the absence of a additional atom layer.
As already mentioned, defected 
the presence of dangling bonds is associated with a higher friction and shorter interfacial distances, indicating 
the presence of many chemical bonds that continuously form and break across the interface.

Similar findings were reported by Kuwahara et al.
in Ref. \cite{PhysRevMaterials.2.073606}. They performed semiempirical simulations of diamond on diamond contacts analyzing friction and wear at various humidity levels. The water molecules intercalated at the interface split under tribological conditions allowing the formation of ether bonds (C-O-C) across the interface, which generates adhesive friction. This result is qualitatively equivalent to the Si-O-C bonds formation observed in dry conditions during the AIMDs presented in this work. In further agreement with our results, water passivation and Pandey reconstruction well lubricate the diamonds interface reducing the resistive force to a value as low as 0.1 GPa. They observed that incrementing the interfacial water content, up to having a water film formation, increases friction. Superficial hydrophilic C-OH groups act like anchors connecting the diamond surfaces through the formation of an H-bonds network with the solvent film. We may expect that H-passivation or Pandey reconstruction can prevent the formation of such H-bonds networks, so maintaining excellent lubricant performances also in wet conditions \cite{DeBarrosBouchet2012}.

It has been shown that the sliding direction of the countersurface strongly influences the amount and rate of diamond-on-diamond wear \cite{nat-mat-mos}. The friction coefficient has similar anisotropic behaviour \cite{Grillo_2000,carpick2007}. Here we analyze this aspect also for the silica-diamond interface. We consider C(111), C(110) and C(001) diamond substrates and move the silica countersurface in perpendicular direction with respect to the previously discussed cases. The simulations results, gathered in Fig. 2 of supporting information, indicate that: i) on the C(111) surface, the resistive force is equivalent in all the considered sliding directions; ii) on the C(001) surface, sliding in the $<$110$>$ direction generates higher resistive force than in the $<$$\bar{1}$10$>$, which is observed also in diamond-on-diamond contacts \cite{carpick2007}; iii) on the C(110) surface, a regime with high resistive force is established more rapidly sliding in the $<$001$>$ direction than in the $<$1$\bar{1}$0$>$ direction.

\subsection{Kinetic of Diamond Surface Passivation}
\begin{figure*}
    \centering
    \includegraphics[width=0.98\textwidth]{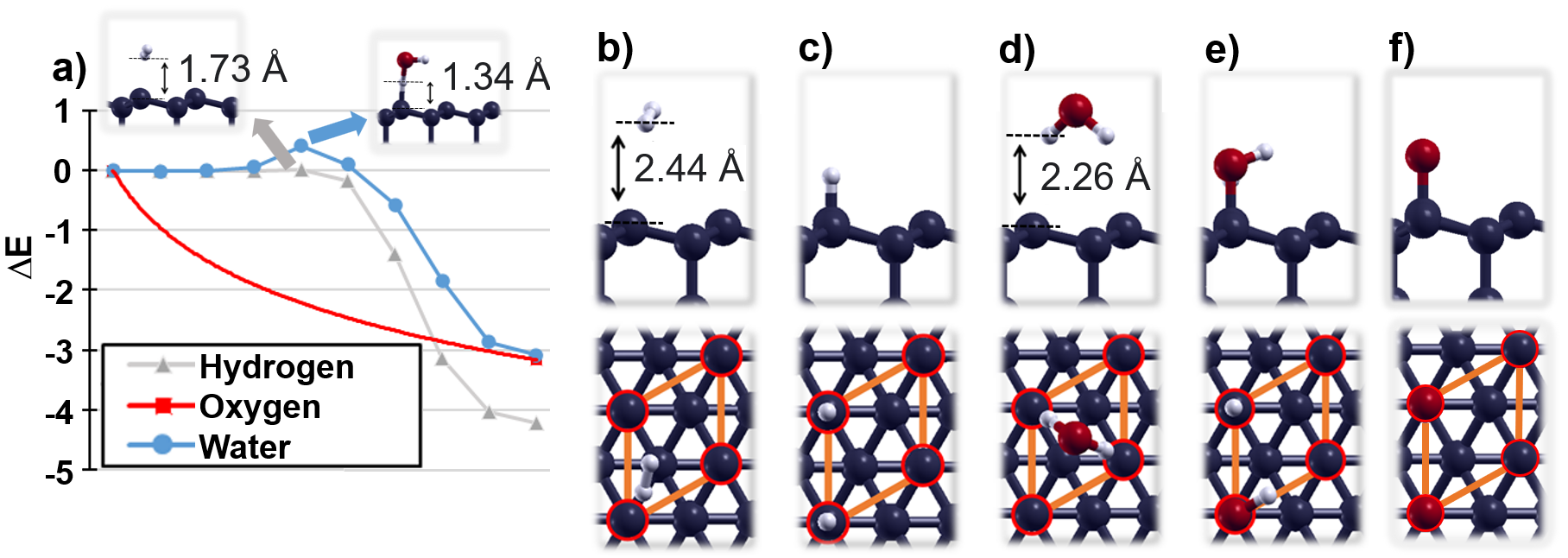}
    \caption{a) Minimum energy paths (in eV) for the single-molecule chemisorption process on the C$\cdot$(111) surface. 
    b)-d): Physisorption geometry of H$_{2}$ and H$_{2}$O species. 
    c)-e)-f): Chemisorption geometry for H$_{2}$, H$_{2}$O and O$_{2}$ species. 
    The distances are evaluated between the most exposed C atom of diamond and the closest atom of the molecule.}
    \label{fig:ts1}
\end{figure*}

This section focuses on the kinetics of diamond surface passivation. Firstly we explore the potential energy
surface associated with ideal single-molecule events on the wholly defective C$\cdot$(111) surface. 
We consider the same molecular species of previous sections of the paper, i.e. H$_{2}$, O$_{2}$ 
and H$_{2}$O, and employ the NEB method to estimate the transition state geometry and energy of the minimum 
energy paths linking physisorbed and chemisorbed molecules. Results are summarized in Fig.~\ref{fig:ts1}.
In ideal conditions, a single O$_{2}$ molecule is unstable on the C$\cdot$(111) surface. It chemisorbs 
spontaneously during the geometry optimization with no energy barrier releasing $3.17\,\text{eV}$ of energy, see Fig.~\ref{fig:ts1}. 
H$_{2}$ and H$_{2}$O are less reactive and can adhere to diamond without causing the molecule to split.
The computed interaction energies, i.e., 0.06 and 0.19 eV, indicate they physisorb on the diamond surface
with optimized geometries shown in panels (b) and (d) of Fig.~\ref{fig:ts1}, respectively. 
The dissociative chemisorption of H$_{2}$ (H$_{2}$O) is not a barrier-less process and occurs with activation energy 
of $0.06\,\text{eV}$ ($0.41\,\text{eV}$) and energy gain of $4.23\,\text{eV}$ ($3.09\,\text{eV}$).

To estimate the surface passivation rate, it is necessary to take into account the probability of the molecule to adsorb/desorb on the surface, which is related to adhesion energy, gas pressure, temperature, as well as the energy needed to the molecule to overcome the transition state energy. Therefore, from the computed energy barriers, it is possible to estimate the time scale of the chemisorption events considering the molecule adhered stably to the surface. This condition is representative of the molecular confinement often occurring at the interface of materials under tribological conditions. Using the Eyring–Polanyi equation~\mbox{\cite{NEB:evans1935,NEB:eyring1935}} we calculated the kinetic constant (k) for the reaction which has as reactant the physisorbed molecule and as product the chemisorbed molecule. This is an unimolecular elementary step of the macroscopic event of gas chemisorption and thus follows a first-order kinetic.\cite{Atkins2008} Clearly, the overall process, involving gas diffusion, adhesion, and chemisorption, may follow a different kinetic model, which is inferable from experimental data \mbox{\cite{elovich}}.
We can easily estimate the half-life time (t$_{1/2}$) of the reaction, which is t$_{1/2}$ = ln(2)/k (first order kinetic). The results indicate that the chemisorption of  H$_{2}$ occurs in the pico-seconds timescale, while that of H$_{2}$O takes place
in the micro-seconds timescale at room temperature.
Regardless the rapidity of these events, the experimental diamond surface is not passivated instantly in contact with H$_2$/H$_2$O/O$_2$-rich atmospheres. Indeed, C(111) surface reconstructs lowering the content of reactive C$\cdot$ defects, and increasing the energy barrier for chemisorption and thus slowing the reaction rates \cite{LEVITA2018533}. A further slowing factor for H$_2$ chemisorption is the low adsorption energy, i.e. 0.06 eV, which is equal to $\sim$ 5/2 kT, at room temperature. So H$_2$ surface adsorption is unlikely to happen, which further prevents the gas chemisorption.

\begin{figure*}
    \centering
    \includegraphics[width=0.98\textwidth]{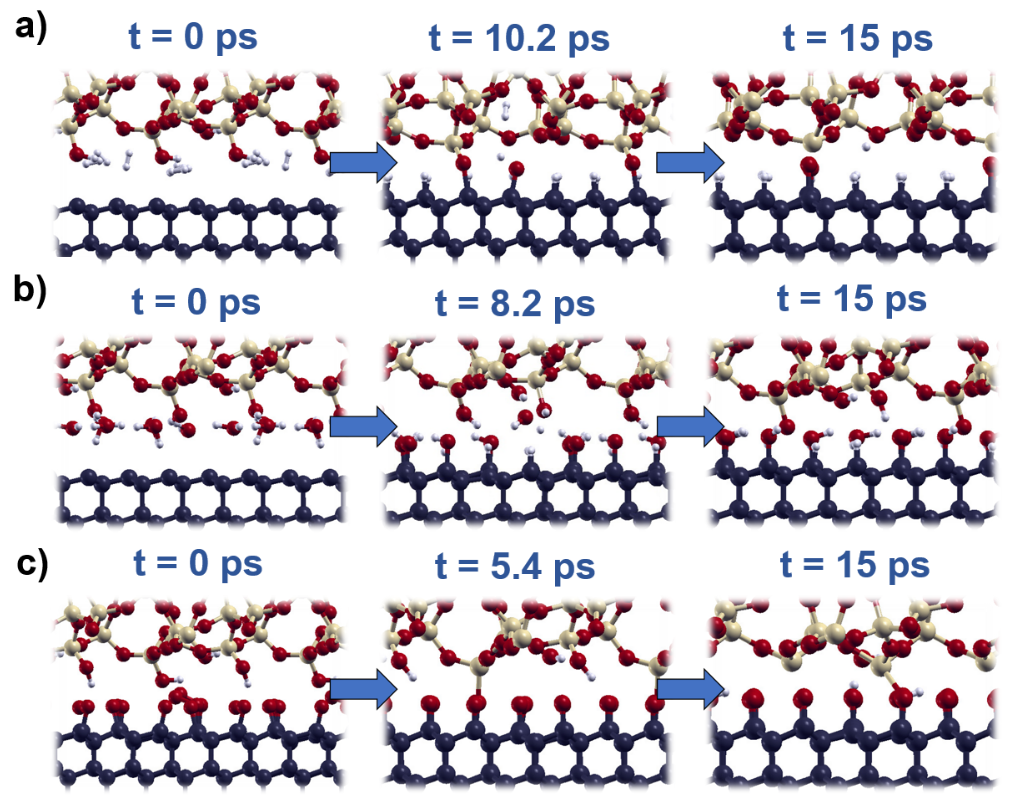}
    \caption{Snapshots of the tribological simulations for the  silica-diamond interface 
    intercalated with  H$_{2}$ (a), H$_{2}$O (b) and O$_{2}$ (c) at 75\% of coverage.}
    \label{fig:ts2}
\end{figure*}

Then we analyze the same process in realistic tribological conditions, which include multiple passivating 
species and a sliding counter-surface under 1GPa of normal load, see Fig. \ref{fig:ts2}. 
We consider two different values for the molecular concentration, i.e., 25\% and 75\% (2 and 6 molecule/cell). 
For the sake of brevity, we discuss only the results for 75\% coverage, relegating those for 25\% coverage to the SI.
Under tribological conditions, already after the initial static relaxation all oxygen molecules, 
but one, have chemisorbed at the initial $t=0\,\text{ps}$ time, see Fig. \ref{fig:ts2} panel a), b) and c). The remaining one chemisorbs during the 
first few steps of the dynamical evolution. This behavior is consistent with the results obtained at the open surface in static
conditions, indicating that the oxygen molecules reacts very quickly with C danging bonds. Results for H$_{2}$ and 
H$_{2}$O cases are, however, somewhat unexpected. Regardless of the small energy barrier, H$_{2}$ turns out
to be slower in passivating the diamond surface, see Fig. \ref{fig:ts2} panel a). 
Its high mobility and low physical attraction by the diamond surface make 
dissociative chemisorption events rare. After 62.5\% of surface passivation is achieved in $10.2\,\text{ps}$, 
no further evolution is observed until the end of the $15\,\text{ps}$ simulations. Instead, H$_{2}$O molecules fast passivate the diamond surface, see Fig. \ref{fig:ts2} panel b). The efficiency of water chemisorption is due to: i) the molecules reduced mobility, which occur by strong H-bonding to silica surface and steric hindrance, and ii) the catalytic effect of silica hydroxyl groups and neighboring water molecules, which open different reaction pathways with lower activation energy. Within $8.2\,\text{ps}$ a passivation level of 62.5\% is achieved and passivation 
is completed (75\% dissociated molecules) within $15\,\text{ps}$.
The so-formed partially passivated surface layer prevents the adhesion of the diamond to the silica countersurface 
during the dynamics. As a result, the tribological figures-of-merit extracted from the AIMD place this case 
in between the defective C$\cdot$(111) surface and fully passivated ones.

\section{Conclusions}

We performed \emph{in silico} tribological experiments for the silica-diamond interface considering a wide number of factors 
relevant to friction and wear. The simulations are carried out with realistic surface models and state-of-the-art first principle methods to ensure highly accurate results by
intensively exploiting HPC resources. Our goal is to provide a general understanding on the atomistic processes determining friction 
and wear at this relevant technological interface, which are not observable directly by experiments, while computational studies have been 
so far limited to specific cases using simplified models and heavily approximated methodology.

The results of our experiments indicate that dangling bonds on diamond surface are primary sources of friction. 
They catalyze the formation of interfacial chemical bonds between the silica and diamond surfaces. An efficient way to reduce the adhesion 
and friction of silicon oxide on diamond is by reducing defects density by hydrogenation. 
All the considered gaseous species, 
i.e., H$_{2}$, O$_{2}$ and H$_{2}$O, have efficient dissociation rates on the unreconstructed C(111) surface and can successfully lower
its friction during sliding against silica, with lubricating efficiency depending more on the passivation level than on the chemical nature 
of the passivating species: The higher the surface coverage, the better the lubrication. The clean C(110) and C(001) surfaces, 
exhibiting graphitic C=C bonds, show less adhesive friction than the unreconstructed C(111) surface, but they do not prevent 
the formation of interfacial bonds under tribological conditions. Conversely, graphitization occurring at the C(111) surface, 
i.e. the Pandey reconstruction, makes the diamond surface inert and has a lubricating capability comparable to full passivation. Further analysis indicates that the silica sliding direction affect the resistive force, in line with previous literature findings.
Silica-driven wear of diamond has been observed only on (110) facet during our simulations, the C(110) surface being
the one with lower stability than those investigated in this work, a part from the unreconstructed C(111) surface, which is reactive enough to wear the silica. Our results clearly indicate that, among the investigated diamond surfaces, the (110) is the one more easily wearable, confirming previous experimental,\cite{Wilks_1972} and computational findings\cite{nat-mat-mos}.
Clearly, longer simulation time would be needed to observe wear events on the other surface orientations, where the energy cost for C detachment is higher. \cite{Thomas2014,Thomas2014a}.

We observe oxygen atom transfer from silica to diamond C(110), which form stable C=O double-bond. Over longer times, 
the observed C=O group may further oxidize, leading to carbon dioxide formation, a standard product of diamond CMP. 
Hydrogen passivation demonstrated to protect the C(110) and C(001) surfaces from wear, even at low hydrogen coverage.
Hydrogen passivation of the diamond is, thus, found to be the most effective approach to reduce friction and wear, being effective 
for all considered diamond surface orientations. 
Tribochemical reactions involving $H_2$, $O_2$ and $H_2O$ molecules have been monitored in real time at the C$\cdot$(111)-silica interface. While oxygen spontaneously dissociate on the diamond reactive surface, the dissociation of hydrogen and water molecules is promoted within the simulation time by the load in shear applied. Interestingly, the rate for molecular dissociation observed for water is higher than for hydrogen, while the dissociation barriers calculated at the open surface in static conditions would have suggested an opposite trend. This suggests that the molecular confinement at the interface influences the reactions rates. Factors, such as the steric hindrance of the molecule, and the presence of catalytic species, should be considered in addition to the energy barrier for surface dissociation when evaluating the efficiency of passivating species.
We believe that the insights presented in this work can help the engineering of better low-friction and wear diamond-silica interfaces that are relevant for many important technological applications.

\section{Acknowledgements}
These results are part of the ``Advancing Solid Interface and Lubricants by First Principles Material Design (SLIDE)'' project 
that has received funding from the European Research Council (ERC) under the European Union's Horizon 2020 research and innovation program 
(Grant agreement No. 865633). Furthermore, we acknowledge PRACE for awarding us access to Marconi100 at CINECA, Italy.

\bibliography{sn-bibliography.bib}%

\end{document}